\documentclass[11pt]{article}
\usepackage{latexsym}
\begin{document}

\newtheorem{theorem}{Theorem}[section]
\newtheorem{lemma}[theorem]{Lemma}
\newtheorem{defin}[theorem]{Definition}
\newtheorem{rem}[theorem]{Remark}
\newtheorem{cor}[theorem]{Corollary}
\newtheorem{prop}[theorem]{Proposition}

 %
 %
 %
 %
\def\a{\alpha}
\def\b{\beta} 
\def\g{\gamma}
\def\d{\delta}
\def\e{\epsilon} 
\def\l{\lambda} 
\def\m{\mu} 
\def\r{\rho}
\def\O{\mbox{\rm O}}

\def\be{\begin{equation}}
\def\ee{\end{equation}}
\def\bea{\begin{eqnarray}}
\def\eea{\end{eqnarray}}
\def\beas{\begin{eqnarray*}}
\def\eeas{\end{eqnarray*}}

\def\dx{\partial_x}
\def\dv{ \partial_v }

\def\N{{\rm I\kern-.1567em N}}
\def\R{{\rm I\kern-.1567em R}}

\newcommand{\prf}{\noindent
         {\em Proof :\ }}

\newcommand{\prfe}{\hspace*{\fill} $\Box$ 

\smallskip \noindent}

\def\supp{\mbox{\rm supp}} 

\def\n#1{\vert #1 \vert}
\def\nn#1{\Vert #1 \Vert}
\def\emin{E_{\rm min}}

\title{Compact support
       of spherically symmetric equilibria in 
       non-relativistic and relativistic galactic dynamics}
\author{Gerhard Rein\\
        Mathematisches Institut der Universit\"at M\"unchen\\
        Theresienstr.\ 39, 80333 M\"unchen, Germany\\
        \ \\
        Alan D.~Rendall\\
        Max-Planck-Institut f\"ur Gravitationsphysik\\
        Schlaatzweg 1, 14473 Potsdam, Germany}
\date{}
\maketitle
\begin{abstract}
Equilibrium states in galactic dynamics can be described as stationary 
solutions of the Vlasov-Poisson system,
which is the non-relativistic case, or of the Vlasov-Einstein system,
which is the relativistic case.
To obtain spherically symmetric
stationary solutions the distribution function of the
particles (stars) on phase space is taken to be a function
$\Phi (E,L)$ of the particle energy and angular momentum. We give a new 
condition on $\Phi$ which guarantees that the resulting
steady state has finite mass and compact support both for the
non-relativistic and the relativistic case. The condition is local in 
the sense that only the asymptotic behaviour of $\Phi$
for $E \to E_0$ needs to be prescribed, where $E_0$ is a cut-off energy
above which no particles exist.  
\end{abstract}
 %
 %
 %
 %
\section{Introduction}
\setcounter{equation}{0}
Let $f=f(t,x,v),\ t \in \R,\ x, v \in \R^3$ be the density function on
phase space of the stars in a galaxy. We assume that collisions
among the stars are sufficiently rare to be neglected and that the
stars interact only by the gravitational field which they
create collectively. In a non-relativistic situation $f$ obeys the 
Vlasov-Poisson system, in a general relativistic situation one obtains the
Vlasov-Einstein system. We are interested in spherically symmetric
equilibrium solutions of these systems, i.~e., $f$ is independent of time $t$,
and
$f(Ax,Av)=f(x,v),\ A \in \mbox{SO}\,(3),\ x, v \in \R^3$.
The Vlasov-Poisson system then takes the form
\be \label{vlasov}
v \cdot \nabla_x f - U' \frac{x}{r} \cdot \nabla _v f=0,
\ee
\be \label{poisson}
U' (r) =\frac{4 \pi}{r^2} \int_0^r s^2 \r (s)\, ds, 
\ee
\be \label{rhodef}
\r (r) =\r (x) = \int f(x,v)\,dv. 
\ee
Here $r=\n{x}$, $'$ denotes derivative with respect to $r$,
$U$ denotes the gravitational potential of the system
and $\r$ the spatial mass density induced by $f$; we assume that 
all the stars have mass one and note that due to spherical symmetry
$U$ and $\r$ depend only on $r$. 
The Vlasov-Einstein system takes the form
\be \label{rvlasov}
\frac{v}{\sqrt{1+v^2}} \cdot \nabla_x f - \sqrt{1+v^2} \m'
\frac{x}{r} \cdot \nabla_v f=0, 
\ee
\be \label{rfield1}
e^{-2\l} \left(2r \l'-1\right) +1 = 8 \pi r^2 \r,
\ee
\be \label{rfield2}
e^{-2\l} \left( 2r \m'+1\right) -1 = 8 \pi r^2 p, 
\ee
\be \label{rrhodef}
\r(r) = \r(x) = \int \sqrt{1+v^2}\ f(x,v) dv,
\ee
\be \label{rpdef}
p(r) = p(x) = \int \left( \frac{x\cdot v}{r} \right)^2 f(x,v)
\frac{dv}{\sqrt{1+v^2}}. 
\ee
Here $v^2 = v\cdot v$, and $\cdot$ denotes the Euclidean scalar
product on $\R^3$.
If $x = \left( r \sin \theta \cos \phi , r \sin \theta \sin
\phi, r\cos \theta\right)$ then the space time metric is given by 
\[
ds^2 = - e^{2\m}dt^2 +e^{2\l} dr^2 + r^2 \left(  d\theta^2 +\sin^2
\theta d \phi ^2\right),  
\]
$\r$ denotes the mass density and $p$ the radial pressure. 
As to the choice of coordinates on phase space, which leads to
the above form of the static, spherically symmetric Vlasov-Einstein
system, we refer to \cite{RR1}. 
As boundary conditions we require asymptotic flatness, i.~e., 
\be \label{asflat}
\lim_{r\rightarrow\infty} \m (r) = 
\lim_{r\rightarrow \infty} \l (r) = 0,
\ee
and a regular center, i.~e., 
\be \label{regcenter}
\l (0) = 0. 
\ee
For the Vlasov-Poisson system we require
\be \label{vpbc}
\lim_{r\rightarrow \infty} U(r)=0, 
\ee
and we will also need the radial pressure $p$ in the 
non-relativistic case, which is defined as
\be \label{pdef}
p(r) = p(x) = \int \left( \frac{x\cdot v}{r} \right)^2 f(x,v)\,dv. 
\ee
There are essentially two approaches to construct solutions
of these systems. The first is to observe that there exist
invariants of the particle motion, namely the particle energy 
$E=E(x,v)$ and angular momentum squared $L=L(x,v)$; 
for the definition of these quantities 
cf.~(\ref{cons}) and (\ref{rcons}) respectively.
The ansatz 
\be \label{ansatz}
f(x,v) = \Phi (E,L)
\ee
with some prescribed function $\Phi$ automatically satisfies the Vlasov
equation, and it remains to solve the field equation(s) with the ansatz
for $f$ substituted into the definitions for $\r$ and $p$; these quantities
become functionals of $U$ or $\m$ since $E$ depends
on $U$ or $\m$ respectively. The main problem then
is to show that the resulting steady state has finite (ADM) mass and
compact support. In \cite{BFH} this was done in the non-relativistic case
for the so-called polytropic ansatz
\[
\Phi (E,L) = (E_0-E)_+^k L^l,
\]
where $E_0$ is some constant, $(\cdot)_+$ denotes the
positive part, and $k>-1,\ l>-1,\ k+l+1/2 >0,\ k< 3l +7/2$.
In \cite{R1,RR3} an existence result was established for the Vlasov-Einstein 
system, exploiting the fact that the Vlasov-Poisson system is the limit of the
Vlasov-Einstein system as the speed of light tends to infinity,
cf.\ \cite{RR2}. (In the present paper the speed of light is 
set to unity.) This perturbation argument does not give good control
over the class of models obtained, a fact which motivates the search for
a better method. It should be noted that in order to obtain a steady state 
with finite mass $\Phi$ must vanish for energy values larger than some
cut-off energy $E_0$.

The second approach is to define an energy-Casimir functional 
which has the property that its
critical points are steady states, and then show that this functional
has a minimizer over a certain set of phase space densities $f$. 
This approach was used in \cite{GR} for the
Vlasov-Poisson system. It has the advantage that it provides a
certain nonlinear dynamical stability property of the steady state obtained,
and the resulting steady states are more general than the polytropic
ones: Only certain growth conditions on $\Phi$ need to be prescribed. 
Nevertheless, the latter approach also makes use of global properties
of the function $\Phi$, i.\ e., of growth conditions 
for $E$ close to $E_0$ and close to $-\infty$. 

The present paper follows the first approach, 
but we give a new characterization of $\Phi$'s which lead to
finite mass and compact support both in the non-relativistic
and in the relativistic case: Except for some
mild technical assumptions we only require that  
\[
\Phi(E,L) = c (E_0 - E)_+^k L^l + \O ((E_0-E)_+^{k+\delta})
L^l \ \mbox{as}\ E \to E_0
\]
where 
\[
k>-1,\ l>-\frac{1}{2},\ k+l+\frac{1}{2} > 0,\
k<l+\frac{3}{2} .
\]
Thus the characterization is purely local:
Only the asymptotic behaviour at $E=E_0$ needs to be controlled.
Since such steady states together with their features of
finite mass and compact support persist under perturbations 
of the ansatz function $\Phi$ as long as the form of the
asymptotic expansion at $E_0$ is preserved, they might be
called structurally stable. Another important point is that we obtain 
finite ADM mass and compact support in the relativistic case directly, 
i.~e., without a perturbation argument as was used in \cite{R1,RR3}.
The assumptions are much more transparent than the smallness conditions
required when starting from the Newtonian limit.

The main idea of the argument is as follows: If $R \in ]0,\infty]$
denotes the radius of the support of a steady state then one can show that
$E_0 - U(r) \to 0$ or
$E_0 - e^{\m (r)} \to 0$ as $r \to R-$. One needs to show that $R<\infty$,
and this is done by expanding  
all the relevant quantities is terms of  $E_0 - U(r)$ or
$E_0 - e^{\m (r)}$, obtaining detailed estimates for
the behaviour of the solution of the field equations as $r \to R-$.
This approach was strongly motivated by the paper of Makino\cite{M} where 
a technique of this kind is used to prove finite mass
and compact support for relativistic,
spherically symmetric stellar models, i.~e., for steady states
of the Euler-Einstein system. The connection between the two situations
is as follows. If a spherically symmetric steady state solution of the 
Vlasov-Einstein system corresponds to a choice of $\Phi$ which only depends 
on $E$, then the energy density $\rho$ and the pressure $p$ define a
solution of the Euler-Einstein system describing a self-gravitating perfect
fluid. (A similar relation between kinetic and fluid models holds in the
non-relativistic case.) The asymptotic behaviour of $\Phi$ which is
important in our results corresponds to the way in which $p$ depends on
$\rho$ (equation of state) in the limit $\rho\to 0$. 

Our paper proceeds as follows: In the next section we 
derive the reduced problems which are obtained by substituting
the ansatz for $f$ into $\r$ and $p$ in the field equation(s).  
We also show that a cut-off energy $E_0$ is necessary in order
to obtain finite mass and compact support. This shows
that except for the form of the dependence on $L$ our ansatz
is quite general.
The main result is then stated and proven in Section 3.
In a last section we consider steady states which appear
in the astrophysics literature and show that our result
applies to most of these and proves 
that these steady states have the physically very
desirable properties of finite mass and compact support.
Clearly, there are some polytropic steady states with
finite mass and compact support which are not covered by
our result: For the polytropes one needs $k< 3 l + 7/2$
whereas we require $k< l + 3/2$. In the last section we 
also comment on  this discrepancy from the viewpoint of the relation
to static self-gravitating fluid bodies and from the viewpoint
of recent stability results for the Vlasov-Poisson system;
cf.~\cite{G,GR}.

To conclude this introduction we mention some further references
which are related to the present paper. Steady states
of the Vlasov-Poisson system with axial symmetry are
constructed in \cite{R2}. Spherically symmetric steady states
with a vacuum region at the center are constructed in
\cite{R3} both for the Vlasov-Poisson and the Vlasov-Einstein system.
Among these there are examples of relativistic
steady states violating
Jeans' Theorem, which holds for the Vlasov-Poisson system
and says that  all
spherically symmetric steady states are obtained by an ansatz of the 
form (\ref{ansatz}), cf.~\cite{Sch2}. In \cite{R4} steady states of the Vlasov-Poisson system are constructed where the matter is concentrated
on a plane, using the variational approach mentioned earlier. 
Concerning the initial value problem for the time dependent
systems we mention \cite{LP,P,Sch1}, where global existence
of classical solutions to the Vlasov-Poisson system is established
for general data. 
For the Vlasov-Einstein system
the existence theory for the initial value problem is far 
less complete. We mention \cite{RR1,RRS} for the spherically 
symmetric, asymptotically flat case which is of interest here.

\section{The reduced problem}
\setcounter{equation}{0}
Consider first the Vlasov-Poisson system. The particle energy
and angular momentum squared 
\be \label{cons}
E=E(x,v)=\frac{1}{2} v^2 + U(r),\ L=L(x,v)= \n{x\times v}^2
\ee
are conserved along particle trajectories, 
i.~e., are constant along solutions of the characteristic system
\[
\dot x = v,\ \dot v = - U'(r) \frac{x}{r}
\]
of (\ref{vlasov}); $r = \n{x}$.
If we make the ansatz (\ref{ansatz})
then upon substituting (\ref{ansatz}) and (\ref{cons})
into (\ref{rhodef}), $\r$ becomes a functional of
$U$, and it remains to solve the resulting nonlinear Poisson equation;
for the moment we only require $\Phi$ to be a nonnegative, measurable function.
Similarly, for the Vlasov-Einstein system one finds that
\be \label{rcons}
E = E(x,v) = e^{\m(r)} \sqrt{1+v^2},\ L = L(x,v)= \n{x\times v}^2
\ee
are constant along solutions of
\[
\dot x = \frac{v}{\sqrt{1+v^2}},\
\dot x = - \sqrt{1+v^2} \m'(r) \frac{x}{r},
\]
and it remains to solve the system (\ref{rfield1}) and (\ref{rfield2})
where $\r$ and $p$ become functionals of $\m$ upon substituting
(\ref{ansatz}) and (\ref{rcons}) into (\ref{rrhodef}) and (\ref{rpdef}). 
It is a simple computation
to see that in the non-relativistic
case substituting (\ref{ansatz}) into (\ref{rhodef}) yields
\be \label{rhodar1}
\r(r)=\frac{2\pi}{r^2}\int_{U(r)}^\infty \int_0^{2r^2(E-U(r))}
\Phi(E,L)\frac{dL\,dE}{\sqrt{2(E-U(r)-L/2r^2)}}.
\ee
In the relativistic case we obtain
\be \label{rrhodar1}
\r(r) =
\frac{2\pi}{r^2} e^{-2\m(r)}\int_{e^{\m(r)}}^\infty
\int_0^{r^2(E^2 e^{-2\m(r)}-1)} 
\Phi(E,L)\frac{E^2\, dL\,dE}{\sqrt{E^2-e^{2\m(r)}(1+L/r^2)}},
\ee
and
\be  \label{rpdar1}
p(r) =
\frac{2\pi}{r^2} e^{-2\m(r)}\int_{e^{\m(r)}}^\infty
\int_0^{r^2(E^2 e^{-2\m(r)}-1)} 
\Phi(E,L)\sqrt{E^2-e^{2\m(r)}(1+L/r^2)}dL\,dE.  
\ee
Before proceeding further we demonstrate that $\Phi$ has to vanish
for large values of $E$ if the resulting steady state is to have 
finite mass, i.~e., if the quantity
\be \label{mdef}
M = \int \r (x)\,dx
\ee
is finite; $M$ is the total mass or the total ADM mass of the steady state respectively. Here and below it will be useful to recall that the solution
of (\ref{rfield1}) satisfying (\ref{regcenter}) 
is given by
\be \label{lambda}
e^{-2\l} =  1-\frac{2 m(r)}{r}
\ee
where
\be \label{mrdef}
m(r) = 4 \pi \int_0^r s^2 \r(s)\, ds,
\ee
at least as long as $2\,m(r) < r$. Using (\ref{lambda}) one can
rewrite (\ref{rfield2}) in the form
\be \label{muprime}
\m'(r) = e^{2\l}\left(\frac{m(r)}{r^2} + 4 \pi r p(r)\right) .
\ee
\begin{theorem} \label{cutoff}
Let $\Phi : \R^2 \to [0,\infty[$ be measurable.
\begin{itemize}
\item[(a)]
Let $(f,U)$ be a solution of (\ref{vlasov}), (\ref{poisson}),
(\ref{rhodef}) in the sense that $f(x,v)=\Phi(E,L)$ and 
$U \in C^1(]0,\infty[)$
solves (\ref{poisson}) with (\ref{rhodar1}) substituted in. 
Let $M < \infty$. 
Then $U_\infty = \lim_{r\to \infty} U (r) < \infty$, and
$\Phi (E,L) = 0$ a.~e.\ for $E > U_\infty,\ L > 0$.
\item[(b)]
Let $(f,\l,\m)$ be a solution of (\ref{rvlasov}), (\ref{rfield1}),
(\ref{rfield2}), (\ref{rrhodef}), (\ref{rpdef}) in the sense that $f(x,v)=\Phi(E,L)$ and $\l,\m \in C^1(]0,\infty[)$
solve (\ref{rfield1}), (\ref{rfield2}) 
with (\ref{rrhodar1}), (\ref{rpdar1}) substituted in.  
Let $M < \infty$. 
Then $\m_\infty = \lim_{r\to \infty} \m (r) < \infty$, and
$\Phi (E,L) = 0$ a.~e.\ for $E > e^{\m_\infty},\ L > 0$.
\end{itemize}
\end{theorem}
\prf
Let us first consider the non-relativistic case. Since 
\[
0 \leq U'(r) \leq \frac{M}{r^2},\ r > 0,
\]
$U$ is increasing and has a finite limit as $r \to \infty$. 
By (\ref{rhodar1}),
\beas
M
&=&
8 \pi^2 \int_0^\infty \int_{U(r)}^\infty \int_0^{2 r^2(E-U(r))}
\Phi(E,L) \frac{dL\,dE\,dr}{\sqrt{2(E-U(r)-L/2r^2)}}\\
&\geq&
8 \pi^2  \int_{U_\infty}^\infty  \int_0^\infty \Phi(E,L)
\int_{\sqrt{L/(2(E-U_\infty))}}^\infty \frac{dr}{\sqrt{2(E-U(r))}}
dL\,dE .
\eeas
The integral with respect to $r$ in the latter expression is infinite 
for any $E>U_\infty$ and $L>0$, which implies that $\Phi$ 
has to vanish for such arguments.

Let us now consider the relativistic case. Since by (\ref{muprime})
$\m$ is increasing, the limit $\m_\infty = \lim_{r \to \infty}
\m(r) \in ]-\infty,\infty]$ exists. Since  
$m(r) \leq M$ and thus $e^{2\l(r)} \leq 2$ for $r \geq 4 M$, 
and $p \leq \r$, we conclude that
\[
\m'(r) \leq 2 \left(\frac{M}{r^2} + 4 \pi r \r(r)\right),\ r\geq 4 M,
\]
and  
\[
\m(r) \leq \m(4 M) + 2 \int_{4M}^\infty \frac{M}{r^2} dr
+ 4\pi\frac{1}{4 M} \int_{4M}^\infty s^2 \r (s)\,ds < \infty,\ r\geq 4 M
\]
which proves that $\m_\infty < \infty$.
Using (\ref{rrhodar1}) we obtain
\beas
M
&=&
8 \pi^2 \int_0^\infty e^{-2\m(r)}
\int_{e^{\m(r)}}^\infty \int_0^{r^2(E^2e^{-2\m(r)}-1)}
\Phi(E,L) \frac{ E^2\,dL\,dE\,dr}{\sqrt{E^2-e^{2\m(r)}(1+L/r^2)}} \\
&\geq&
8 \pi^2 e^{-2\m_\infty} \int_{e^{\m_\infty}}^\infty  \int_0^\infty 
 E^2 \Phi(E,L)
\int_{\sqrt{L/(E^2 e^{-2\m_\infty}-1)}}^\infty 
\frac{dr}{\sqrt{E^2-e^{2\m(r)}}} 
dL\,dE,
\eeas
and the assertion follows as in the non-relativistic case. \prfe

Although this will become important only in the next section
where we prove our main result we  restrict
the ansatz (\ref{ansatz}) in the following way:
\be \label{ransatz}
f(x,v) = \phi(E) L^l
\ee
with $l>-1/2$ and $\phi$ measurable.  Observe now that in the 
non-relativistic
case $E$ can attain any real value, whereas in the relativistic
case $E>0$. In order not to introduce 
unnecessary assumptions this needs to be reflected in the
choice of domain of $\phi$, which we denote by $]\emin,\infty[$
with
\[
\emin = \left\{
\begin{array}{cl} 
-\infty & \mbox{in the non-relativistic case,}\\
0 & \mbox{in the relativistic case.} 
\end{array}
\right.
\]
Since we are interested only in 
steady states with finite mass we assume that
$\phi (E) = 0$ for all energies $E$ larger than some given $E_0 > \emin$.
It will be necessary to
have some information on the functional dependence of $\r$ and $p$ on
$U$ or $\m$ respectively:
\begin{lemma} \label{regularity}
Let $\phi : ]\emin,\infty[ \to \R$ be measurable, $E_0 > \emin$, 
and $k>-1$ such that 
on every compact subset $K \subset ]\emin,\infty[$ there exists $C\geq 0$
such that
\[
0 \leq \phi (E) \leq C (E_0 - E)_+^k,\ E \in K.
\]
Define
\beas
g_m (u) 
&=& 
\int_u^\infty \phi(E) (E-u)^m dE,\ u\in ]\emin,\infty[,\\
h_m (u)  
&=&
\int_u^\infty  \phi(E) (E^2-u^2)^m dE,\ u\in ]\emin,\infty[.
\eeas
If $m>-1$ and $k+m+1>0$ 
then $g_m, h_m \in C(]\emin,\infty[)$.
If $m>0$ and $k+m>0$ then
$g_m, h_m \in C^1(]\emin,\infty[)$
with 
\[
g'_m = -m\, g_{m-1},\ h'_m = - 2 m\, u\, h_{m-1}.
\]
\end{lemma}
\prf
The continuity assertions can be
obtained using Lebesgue's dominated convergence theorem.
By the same tool one can show that the functions are left differentiable
with the given derivatives. Since these left derivatives
are again continuous the functions are differentiable as claimed.
The details are fairly lengthy but largely technical, and are therefore
omitted. \prfe

Using the lemma above we express $\r$ and $p$ as functionals
of $U$ or $\m$ respectively. To this end we introduce for
$a >-1,\ b > -1$ the constants
\be \label{cab}
c_{a, b} = \int_0^1 s^a (1-s)^b ds = 
\frac{\Gamma(a+1) \Gamma(b+1)}{\Gamma(a + b + 2)},
\ee
where $\Gamma$ denotes the gamma function.
It will be useful later to note that
\be \label{cabfrac}
\frac{c_{a,b-1}}{c_{a,b}} = \frac{a+b+1}{b},\ a>-1,\ b>0,
\ee
which follows from the functional relation $x \Gamma(x) = \Gamma(x+1)$.
If $\phi$ is as in the lemma above and $f$ given by the ansatz (\ref{ransatz}) then in the non-relativistic case we obtain
\bea \label{rhodar2}
\r (r) 
&=&
2^{l+3/2} \pi c_{l,-1/2} r^{2l} g_{l+1/2} (U(r)),\\
\label{pdar2}
p(r)
&=&
2^{l+5/2} \pi c_{l,1/2} r^{2l} g_{l+3/2} (U(r)).
\eea
For the relativistic case we substitute $E^2 = (E^2-e^{2\m}) + e^{2\m}$
into (\ref{rrhodar1}) and use the abbreviation
\be \label{h}
H_l(e^\mu)
= e^{-(2l+4)\m} h_{l+3/2}(e^{\m})
+  e^{-(2l+2)\m} h_{l+1/2}(e^{\m})
\ee
to obtain
\bea \label{rrhodar2}
\r (r) 
&=&
2 \pi c_{l,-1/2} r^{2l} H_l (e^{\m(r)}), \\ 
\label{rpdar2}
p(r)
&=&
2 \pi c_{l,1/2} r^{2l} e^{-(2l+4)\m(r)} h_{l+3/2}(e^{\m(r)}).
\eea
A steady state of the Vlasov-Poisson
system is now obtained by solving the equation
\be \label{redpoisson}
U' (r) = \frac{2^{l+7/2} \pi^2 c_{l,-1/2}}{r^2} \int_0^r 
 s^{2l+2} g_{l+1/2} (U(s))\, ds,\ r>0,
\ee
which is (\ref{poisson}) with (\ref{rhodar2}) substituted in,
in the case of the Vlasov-Einstein system one needs to solve
\bea
e^{-2\l} (2 r \l' -1) +1 
&=& 
16\pi^2 c_{l,-1/2} r^{2l+2} H_l(e^\m),\label{rredfield1}\\
e^{-2\l} (2 r \m' +1) -1 
&=& 
16 \pi^2 c_{l,1/2} r^{2l+2} e^{-(2l+4)\m} h_{l+3/2}(e^\m),
\label{rredfield2}
\eea
which is (\ref{rfield1}), (\ref{rfield2}) with
(\ref{rrhodar2}) and (\ref{rpdar2})  substituted in.
It is known that these problems have solutions on $[0,\infty[$,
and we state the corresponding result for further reference;
the problem of interest of course is whether these solutions
lead to steady states with compact support and finite mass.
\begin{theorem} \label{glex}
Let $\phi$ be as in Lemma~\ref{regularity}.
\begin{itemize}
\item[(a)]
Let $U_0 \in \R$. Then there exists a unique solution 
$U \in C^1 ([0,\infty[)$ of (\ref{redpoisson}) with $U(0)=U_0$.
\item[(b)]
Let $\m_0 \in \R$. Then there exists a unique solution 
$(\l,\m) \in C^1([0,\infty[)^2$ of 
(\ref{rredfield1}), (\ref{rredfield2}) with $\l(0)=0$, $\m(0)=\m_0$.
\end{itemize}
\end{theorem}
\prf
In both cases local existence on some interval $[0,\delta]$ follows by a
contraction argument,\ cf. \cite[Thm.~3.6]{BFH} for details in the
non-relativistic case and \cite[Thm.~3.2]{R1} for the relativistic case.
In the non-relativistic case global existence is simple: Clearly, $U$ is
increasing. Either $U \leq E_0$ on its maximal interval of existence,
in which case $U$ exists globally in $r$, or $U(r) > E_0,\ r\geq R$
for some $R>0$, in which case $\r (r)=0,\ r \geq R$, and again $U$ exists
globally. For the relativistic case the
inequality $2 \,m(r) < r/2$ has to be controlled
which makes the argument more involved than in the non-relativistic case, 
cf.\ \cite[Thm.~3.4]{R1}. \prfe 

\noindent
{\bf Remark:} In the theorem above nothing is said about
the boundary conditions at infinity. However, once we know that
a steady state has finite mass then $U$ or $\m$ has a finite limit
at infinity, cf.\ Theorem~\ref{cutoff}. 
Subtracting this limit from $U$ or $\m$
respectively and redefining $E_0$ accordingly gives
a steady state with the same $f$, but which now satisfies the boundary
condition at infinity; the boundary condition for $\l$
follows from (\ref{lambda}) if the ADM mass is finite.
\section{The main result}
\setcounter{equation}{0}
The following theorem is the main result of the present paper:
\begin{theorem} \label{main}
Let $k,l \in \R$ be such that
\[
k>-1,\ l>-\frac{1}{2},\ k+l+\frac{1}{2} > 0,\
k<l+\frac{3}{2} .
\]
Let $\phi : ]\emin,\infty[ \to [0,\infty[$ be measurable and such that
$\phi \in L^\infty_{{\rm loc}}(]\emin,E_0[)$, and
\be \label{phiass}
\phi(E) = c (E_0 - E)_+^k + \O ((E_0-E)_+^{k+\delta}) \ \mbox{as}\ E \to E_0-
\ee
for some $E_0 > \emin,\ c>0$, and $\delta >0$. 
\begin{itemize}
\item[(a)]
Let $(f,U)$ be a steady state of the Vlasov-Poisson system
in the sense that
$f(x,v)=\phi(E) L^l$
with $E$ and $L$ as defined in (\ref{cons}), and 
$U \in C^1([0,\infty[)$ satisfies (\ref{poisson}).
Then the steady state has compact support and finite mass. 
\item[(b)]
Let $(f,\l,\m)$ be a steady state of the Vlasov-Einstein system
in the sense that 
$f(x,v)=\phi(E) L^l$
with $E$ and $L$ as defined in (\ref{rcons}), and 
$\l,\m \in C^1([0,\infty[)$ satisfy (\ref{rfield1}),
(\ref{rfield2}).
Then the steady state has compact support and finite ADM mass. 
\end{itemize}
\end{theorem}
Clearly, if $\phi$ is as in the theorem above then it satisfies the
assumptions in Lemma~\ref{regularity} so that this lemma and 
Theorem~\ref{glex} apply.
The main tool in the proof of Theorem~\ref{main} is the following
lemma, which is an adaptation of \cite[Thm.~1]{M} to our present situation:
\begin{lemma} \label{makino}
Let $x, y \in C^1(]0,R[)$ be such that $x, y > 0$ and
\beas
r x'
&=&
\a (r) y - x + x \frac{x+\g_1(r) y}{1-\g_2(r)x}\\
r y'
&=&
y\, \left( c - \b (r) \frac{x+\g_1(r) y}{1-\g_2(r)x}\right)
\eeas
on $]0,R[$, where $c>0$, $\a,\b,\g_1,\g_2 \in C(]0,R[)$ with
$\a_0 = \inf_{r\in ]0,R[} \a (r) >0$, 
$\lim_{r\to R} \b (r) =\b_0 \in ]0,c[$,
$\g_1, \g_2 \geq 0$, and $\lim_{r\to R} \g_1 (r) =
\lim_{r\to R} \g_2 (r) = 0$.
Also let $1-\g_2(r) x(r) > 0,\ r \in ]0,R[$. Then $R < \infty$.
\end{lemma}
\prf
As a first step we show that there exists $r_\ast \in ]0,R[$ such that
$x(r_\ast) >1$. If not, then $x(r) \leq 1,\ r\in ]0,R[$. By assumption there exists $r_0 \in ]0,R[$ such that $\b(r) >0,\ r\in [r_0,R[$. Choose
$K>0$ such that $K ( 1+x(r_0)) -y(r_0) >0$ and $K \a_0 > 1 + c$. If for some
$r \geq r_0$, $K (1+x(r))-y(r) =0$ then
\beas
r\left(K (1+x) -y \right)'(r)
&=& K r x'(r) - ry'(r)\\
&\geq&
K \a_0 y(r) - K x(r) - c y(r)\\
&=&
(K \a_0 - c - 1) y(r) + K > 0.
\eeas
This implies that no such $r$ exists, and
\[
K (1+x(r)) -y(r) > 0,\ r \in [r_0,R[.
\]
By our assumption $x \leq 1$, and
\[
y(r) \leq K (1+x(r)) \leq 2 K,\ r \in [r_0,R[.
\]
This implies that $R=\infty$, and
\[
r y' \geq
y \left( c -\b(r) \frac{1+ 2 K \g_1(r)}{1-\g_2(r)}\right)
\geq
\frac{c-\b_0}{2} y
\]
for all $r\geq r_1$ sufficiently large, cf.\ the assumption on $\b$.
Integration of this inequality implies
\[
y(r) \geq y(r_1) (r/r_1)^{(c-\b_0)/2} \to \infty,\ r \to \infty,
\]
a contradiction.

Thus we can assume that there exists some $r_\ast \in ]0,R[$
with $x(r_\ast) >1$. Now
\be \label{xprime}
r x' \geq -x + x^2 = x(x-1)
\ee
which implies that $x(r) >1$ for all $r \in [r_\ast,R[$, and upon
integration of (\ref{xprime}),
\[
x(r) \geq \left(1-\frac{x(r_\ast)-1}{x(r_\ast)}\frac{r}{r_\ast}\right)^{-1},\
r \in [r_\ast,R[.
\]
Since the term in parenthesis vanishes for 
$r=r_\ast x(r_\ast)/(x(r_\ast)-1)$ this implies that
\[
R \leq r_\ast \frac{x(r_\ast)}{x(r_\ast)-1}.
\]
\prfe 

\noindent
{\em Proof of Theorem~\ref{main}:}\\ 
{\em Step 1---The basic set-up:}
Consider a solution of the reduced field equation(s) as given by
Theorem~\ref{glex}. Consider the non-relativistic case first 
and define $[0,R[$ as the maximal interval on which
$U < E_0$; we may assume that $U(0)<E_0$, or else the
solution is trivial. If $R<\infty$ then $U(R)=E_0$. If $R=\infty$
then $U_\infty = \lim_{r\to \infty} U(r) \leq E_0$ exists
by the monotonicity of $U$. Assume $U_\infty < E_0$. Then
(\ref{rhodar2}) and the monotonicity of $g_{l+1/2}$ imply that 
\[
\r(r) \geq 2^{l+3/2} \pi c_{l,-1/2} r^{2l} g_{l+1/2} (U_\infty)
= c r^{2l},\ r>0,
\]
where $c>0$. Then (\ref{poisson}) implies that
$U'(r) \geq c r^{1+2l},\ r>0,$ with a different positive constant $c$,
and integrating this estimate implies that $U_\infty = \infty$,
a contradiction. Thus we have the following \\
{\em Basic set-up:}
There exists some $R \in ]0,\infty]$ such that
$U$ exists on $[0,R[$ with $U<E_0$ on this interval, and
\[
\lim_{r\to R-} U(r) = E_0.
\]
Analogously, we have for the relativistic case the\\
{\em Basic set-up:}
There exists some $R \in ]0,\infty]$ such that
$\l,\m$ exist on $[0,R[$ with
$e^\m < E_0$ on this interval, and
\[
\lim_{r\to R-} e^{\m(r)} = E_0.
\]
We may assume that 
$e^{\m (0)} < E_0$, or else the solution is trivial.
We choose $R$ maximal such that $e^\m < E_0$ on $[0,R[$.
The non-obvious case is $R=\infty$. By monotonicity,
$\m_\infty = \lim_{r\to R-} \m(r) \leq E_0$ exists.
Assume that $\m_\infty < E_0$. Then again
$\r(r) \geq c r^{2l},\ r>0$,
with a positive constant $c>0$. 
By (\ref{lambda}) and (\ref{muprime}),
\beas
\m'(r) 
&=&
 \left(1-\frac{8\pi}{r}\int_0^r s^2\r(s)\, ds\right)^{-1}
\left( \frac{4\pi}{r^2}\int_0^r s^2\r(s)\,ds + 4 \pi r p(r) \right)\\
&\geq&
\frac{c}{r^2}\int_0^r s^{2+2l}ds = cr^{1+2l},\ r>0.
\eeas
Integration of this inequality implies that $\m_\infty = \infty$,
a contradiction.

What we need to show in both cases is that $R<\infty$.

\noindent
{\em Step 2---New variables:}
We introduce new variables which bring the 
system into the form stated in Lemma~\ref{makino}.
We define for the non-relativistic case
\be
\eta (r) = E_0 - U(r),
\ee 
and for the relativistic case
\be
\eta (r) = \ln E_0 - \m (r);
\ee
recall that $E_0 >0$ in the relativistic case. 
Now define in both cases
\bea
x(r) 
&=&
\frac{m(r)}{r \eta (r)} = \frac{4 \pi}{r \eta (r)} \int_0^r s^2\r(s)\,ds,\\
y(r)
&=&
4 \pi r^2 \frac{\r^2(r)}{p(r)}
\eea
on the interval $]0,R[$ with $R$ from the previous step;
note that $\eta, \r, p >0$ on that interval.
In the non-relativistic case,
$r\eta' = - r U' = - m/r$, 
whereas in the relativistic case by (\ref{muprime}), 
\[
r \eta' = - r \m' = - \frac{\eta x + y p^2/\r^2}{1-2 \eta x} .
\]
Thus in both cases
\be \label{etaprime}
r \eta'= - \eta \frac{x+ \g_1 (r) y}{1-\g_2 (r) x},
\ee
where in the non-relativistic case
\be \label{ngammas}
\g_1 = \g_2 =0,
\ee 
and in the relativistic case
\be \label{rgammas}
\g_1 = \frac{p^2}{\eta \r^2},\ \g_2 = 2 \eta .
\ee
Now  
\[
rx'=\frac{4 \pi r^2 \r}{\eta} - x - x \frac{r\eta'}{\eta}
= \a(r) y - x + x \frac{x+ \g_1 (r) y}{1-\g_2 (r) x},
\]
where  
\[
\a = \frac{p}{\eta \r}.
\]
In the non-relativistic case we find, using
(\ref{rhodar2}), (\ref{pdar2}), and (\ref{cabfrac}), that
\be \label{nalpha}
\a =
\frac{1}{l+3/2}\frac{g_{l+3/2}(U)}{\eta g_{l+1/2}(U)}.
\ee
In the relativistic case we find, using
(\ref{rrhodar2}), (\ref{rpdar2}), and (\ref{cabfrac}), that
\be \label{ralpha}
\a 
=
\frac{1}{2l+3} \frac{h_{l+3/2}(e^\m)}
{\eta h_{l+3/2}(e^\mu) + e^{2\m} \eta h_{l+1/2}(e^\m)}.
\ee
Next consider the equation for $ry'$. In both cases we find
\[
ry'=2y+\frac{2r\r'}{\r} y - \frac{r p'}{p} y.
\]
In the non-relativistic case Lemma~\ref{regularity} and (\ref{rhodar2}),
(\ref{pdar2}) imply
\[
\r' = \frac{2l}{r} \r - (l+1/2) 2^{l+3/2} \pi c_{l,-1/2}
r^{2l} g_{l-1/2}(U)\, U'
\]
and
\be \label{ntov}
p' = \frac{2l}{r} p - \r\, U'.
\ee
With the definition of $\eta$ and (\ref{etaprime}) this implies that
\[
ry'=y \left((2l+2) - \b (r) \frac{x+\g_1(r) y}{1-\g_2(r)x} \right),
\]
where
\be \label{nbeta}
\b = - \frac{\eta \r}{p} + (2l+1) \eta \frac{g_{l-1/2} (U)}{g_{l+1/2}(U)} .
\ee
In the relativistic case Lemma~\ref{regularity} and (\ref{rrhodar2}),
(\ref{rpdar2}) imply 
\[
\r' =
\frac{2l}{r} \r - 2 \pi c_{l,-1/2} r^{2l} 
\widetilde{H}_l (e^\m) \m'
\]
where
\bea 
\widetilde{H}_l (e^\mu)
&=&
(2l+4) e^{-(2l+4)\m} h_{l+3/2} (e^\m) + 
(4l+5) e^{-(2l+2)\m} h_{l+1/2} (e^\m) \nonumber\\ 
&& {}
+ (2l+1) e^{-2l\m} h_{l-1/2} (e^\m) \label{htilde},
\eea
and 
\[
p' = \frac{2l}{r} p - (p + \r)\m',
\]
which is the Tolman-Oppenheimer-Volkov equation.
With the definition of $\eta$ and (\ref{etaprime}) this implies that
\[
ry'=y \left((2l+2) - \b (r) \frac{x+\g_1(r) y}{1-\g_2(r)x} \right),
\]
where
\be \label{rbeta}
\b = - \eta - \frac{\eta \r}{p} + 
2 \eta \frac{\widetilde{H}_l (e^\m)}{H_l (e^\m)} .
\ee
Thus both in the non-relativistic and in the relativistic case we
obtain a system of the form which is stated in Lemma~\ref{makino},
and in the next two steps
we will show that $\g_1,\g_2,\a$, and $\b$ satisfy the necessary assumptions.

\noindent
{\em Step 3---Application of Lemma~\ref{makino}, the non-relativistic case:}
In this case $\g_1 = \g_2 =0$, cf.~(\ref{ngammas}), so these functions
satisfy the assumptions in Lemma~\ref{makino}.
To investigate the asymptotic behaviour of $\a$ and $\b$
we need to use the  
asymptotic expansion of $\phi$; note that by Step 1, $\eta(r) \to 0$
for $r \to R$. First of all we may assume $c=1$ in (\ref{phiass}),
since this factor cancels in $\a$ and $\b$. A simple
computation shows that
\be
\int_u^{E_0} (E_0 - E)^a (E-u)^b dE =
c_{a,b} (E_0 - u)^{a+b+1},\ u\leq E_0,\ a>-1,\ b>-1
\ee
where $c_{a,b}$ is defined by (\ref{cab}).
Thus
\be \label{gmexp}
g_{m} (u) = c_{k,m} \eta^{k+m+1} + \O(\eta^{k+m+\delta +1}),\
\eta = E_0 - u \to 0+.
\ee
Using (\ref{gmexp}) and (\ref{cabfrac}) in
(\ref{nalpha}) we find that
\[
\a(r) = \left( k+l+\frac{5}{2}\right)^{-1} + \O(\eta(r)^\delta ),\
r \to R-.
\]
Since we assume $U(0) < E_0$,
\[
\a (r) \to \frac{1}{l+3/2}
\frac{g_{l+3/2}(U(0))}{(E_0-U(0)) g_{l+1/2}(U(0))} > 0,\ r \to 0+,
\]
and since $\a >0$ on $]0,R[$ we have shown that
\[
\inf_{r\in ]0,R[} \a (r) > 0
\]
as required. Using (\ref{gmexp}) and (\ref{cabfrac}) in
(\ref{nbeta}) we see that
\[
\b (r) = - (k+l+5/2) + (2l+1) \frac{k+l+3/2}{l+1/2} +\O(\eta(r)^\delta)
\to k+l+\frac{1}{2},\ r \to R-,
\]
and by our assumptions
on $k$ and $l$ this limit lies in the interval $]0,2l+2[$ as required. 
Applying Lemma~\ref{makino} 
we find that $R<\infty$, and the proof of the theorem is complete
in the non-relativistic case.

\noindent
{\em Step 4---Application of Lemma~\ref{makino}, the relativistic case:}
By (\ref{rgammas}) and since $\eta (r) \to 0$ as 
$r \to R-$, $\g_2$ is as required.
We use (\ref{phiass}) to compute the asymptotic behaviour
of the various other quantities as $r\to R-$. To this end we first observe that
(\ref{phiass}) implies that
\[
\phi(E) = c' E\,(E_0^2 - E^2)_+^k + E\, 
\O ((E_0^2-E^2)_+^{k+\delta}) \ \mbox{as}\ E \to E_0-,
\]
which is more suitable for computing the integral
$h_m$. Again we assume without loss of generality that $c'=1$.
Using the relation
\[
\int_u^{E_0} E (E_0^2 -E^2)^a (E^2-u^2)^b dE
= \frac{1}{2} c_{a,b} (E_0^2 - u^2)^{a+b+1},\ u\leq E_0,\ a >-1,\ b>-1,
\]
we find that
\bea
h_m(e^\m) 
&=&
\frac{1}{2} c_{k,m} (E_0^2 - e^{2\m})^{k+m+1} + \O((E_0^2 - e^{2\m})^{k+m+\delta+1}) \nonumber\\
&=&
\frac{1}{2} c_{k,m} \e^{k+m+1} + \O(\e^{k+m+\delta+1}),\
\e = E_0^2 - e^{2\m} \to 0+. \label{hmexp}
\eea 
In the following we need to observe that
\be \label{etaeps}
\eta = \frac{1}{2E_0^2} \e + \O(\e^2),\ \e \to 0+
\ee
and
\be \label{enullmu}
E^2_0 e^{-2\m(r)} = 1 + \O(\e(r)),\ r \to R-.
\ee
Using (\ref{hmexp}), (\ref{etaeps}), (\ref{enullmu}), and (\ref{cabfrac})
in (\ref{ralpha}) we find that
\[
\a (r) =
\frac{1}{2l+3} 2 E_0^2 e^{-2 \m (r)} 
\frac{ c_{k,l+3/2}}{ c_{k,l+1/2}} + \O(\e (r)^\delta) =
\frac{1}{k+l+5/2} + \O(\e (r)^\delta),\ r \to R-.
\]
Since $\a >0$ on $]0,R[$ and $\lim_{r\to 0+} \a (r) > 0$
we find that 
\[
\inf_{r\in ]0,R[} \a(r) > 0
\]
as required. 
Since by (\ref{rgammas}) $\g_1 = \a p / \r$, (\ref{rrhodar2})
and (\ref{rpdar2}) together with (\ref{hmexp}) imply that
$\g_1 (r) \to 0$ as $r \to R-$. 

It remains to examine 
the function $\b$. 
Using (\ref{hmexp}) in (\ref{h}) and (\ref{htilde})
we find
\beas
H_l(e^\m)
&=&
\frac{1}{2} c_{k,l+1/2} e^{-(2l+2)\m} \e^{k+l+3/2}
+ \O(\e^{k+l+\d + 3/2}),\\
\widetilde{H}_l (e^\m)
&=&
\frac{1}{2} c_{k,l-1/2} (2l+1) e^{-2l\m} \e^{k+l+1/2}
+ \O(\e^{k+l+\d + 1/2}),
\eeas
and using this together with
(\ref{etaeps}), (\ref{enullmu}), and (\ref{cabfrac}) in (\ref{rbeta})
we obtain 
\beas
\b (r)
&=&
\O(\e(r)) -\a^{-1} (r) + 2(2l+1) \frac{e^{2\m(r)}}{2 E_0^2}
\frac{c_{k,l-1/2}}{c_{k,l+1/2}} + \O(\e^\d(r))\\
&=&
- (k+l+5/2) + 2(k+l+3/2) + \O(\e^\d(r))\\
&\to&
k+l+\frac{1}{2},\ r \to R-,
\eeas
and  by our assumptions
on $k$ and $l$ this limit lies in the required interval $]0,2l+2[$. 
Applying Lemma~\ref{makino} completes the proof in the
relativistic case.  \prfe

\noindent
{\bf Remark:} We note that the application of Lemma~\ref{makino} 
provides an explicit upper bound on the radius $R$
of the spatial support in terms of a point $r_\ast >0$ at which
\[
x(r_\ast) = \frac{m(r_\ast)}{r_\ast \eta(r_\ast)} > 1
\]
namely
\[
R \leq r_\ast \frac{x(r_\ast)}{x(r_\ast) -1} .
\] 
Such a point $r_\ast$ must exist by the proof of Lemma~\ref{makino},
and this upper bound may be useful in numerical work on steady states.

We have shown that the spatial support
of a steady state as in Theorem~\ref{main} is compact, but
a bound on $v$ over the support of $f$ follows by the boundedness
of $E$ from above together with (\ref{cons}) or
(\ref{rcons}) and the boundedness of $U$ or $\m$ respectively.    

\section{Examples and final remarks}
\setcounter{equation}{0}
The following is a list of types of static spherically symmetric models
built of self-gravitating collisionless matter, non-relativistic and
relativistic, to be found in the literature. In this list inessential 
multiplicative constants are omitted.

\begin{itemize}
\item[(NR1)]
Polytropic solutions of Vlasov-Poisson system. The distribution 
function is of the form $(E_0-E)^k L^l$ for $E<E_0$ and zero otherwise. 
For appropriate $k$ and $l$---cf.\ the introduction---the existence of 
solutions with finite radius was 
proved in \cite{BFH}. Theorem~\ref{main} applies, but only 
to a subclass of these, and we comment on this fact below.  

\item[(NR2)]
King models, cf.\ \cite[p.\ 232]{BT}.
The distribution function is 
of the form $e^{E_0-E}-1$ for $E<E_0$ and zero otherwise. Since
$$e^{E_0-E}-1=(E_0-E)+\O((E_0-E)^2),\ E \to E_0-,$$  
Theorem~\ref{main} applies to these models.

\item[(NR3)]
Woolley-Dickens models, cf.\ \cite[p.\ 235]{BT}.
The distribution
function is of the form $e^{E_0-E}$ for $E<E_0$ and zero otherwise. 
Since 
\[
e^{E_0-E}=1+\O((E_0-E)),\ E \to E_0-,
\]  
Theorem~\ref{main} applies.

\item[(NR4)]
Wilson models,  cf.\ \cite[p.\ 235]{BT}. 
The distribution function is 
of the form $e^{E_0-E}-1-(E_0-E)$ for $E<E_0$ and zero otherwise.  
Since 
\[
e^{E_0-E}-1-(E_0-E) = \frac{1}{2} (E_0-E)^2 + \O((E_0 - E)^3),\ E \to E_0-,
\]
this model is outside the range of the present approach.

\item[(R1)]
Truncated Maxwell-Boltzmann models, cf.\ \cite[p.\ 59]{ST}.
The
distribution function is given by an expression formally identical to that
of the Woolley-Dickens models (of course $E$ has a different definition in 
the two cases), and Theorem~\ref{main} applies.

\item[(R2)]
Power-law models,  cf.\ \cite[p.\ 68]{ST}.
The distribution function 
is given by $[(E/E_0)^2]^{-\delta}[1-(E/E_0)^2]^\delta$
for $E<E_0$ and zero otherwise.  
Since
\[
[(E/E_0)^2]^{-\delta}[1-(E/E_0)^2]^\delta
= \frac{2^\d}{E_0^\d} (E_0 - E)^\d + \O((E_0-E)^{2\d}),\ E \to E_0-,
\]
Theorem~\ref{main} applies if $-1/2 < \d < 3/2$.
\item[(R3)]
Polytropic models, cf.\ \cite[p.\ 68]{ST}, \cite{F}. 
The idea here is the following. Polytropic solutions of the Vlasov-Poisson 
system in the sense of (NR1) with $l=0$ correspond to self-gravitating fluid 
models with $p=\r^{(n+1)/n}$, where $\r$ is the density and $p$ is the 
pressure. This is due to the relation between kinetic and fluid models
mentioned briefly in the introduction and discussed in more detail below.
In general relativity it is also possible to look for a collisionless
model with a given equation of state but there is no unique generalization
of the polytropic case. In \cite{T} Tooper considered different 
possibilities and their relationships. One possibility is
to consider the equation of state which is formally identical to the 
non-relativistic polytropic one with $\r$ being interpreted as the energy 
density.
If, on the other hand, $\r$ is interpreted as the mass density a 
one-parameter family of relations between pressure and energy density 
is obtained. Some authors, e.g. Fackerell\cite{F}, have considered the problem 
of producing corresponding distribution functions. Numerical calculations
indicate that this is sometimes possible but not always. We can prove the
relative statement that when it is possible (and the polytropic index $n$
is restricted appropriately) the radius of the configuration is finite.
Note that the in the case where $\Phi$ depends on $E$ alone the asymptotic
behaviour of $\Phi$ goes into the proof of Theorem 3.1 only via the 
asymptotic behaviour of the equation of state of the corresponding fluid 
model. In the first case the equation of state is directly in the correct 
form. In the second case the relation between pressure and energy density is
$\r=Cp^{n/(n+1)}+np$ for a positive constant $C$. Then for small $\r$
$$p=C^{-(n+1)/n}\r^{(n+1)/n}(1+\r^{1/n})$$

\end{itemize}

Next we want to state two observations which may help to understand 
the relevance of the upper bound $k< l+3/2$ that we required in
our ansatz. We restrict ourselves to the non-relativistic case.
The first observation is based on the correspondence of steady states of 
the Vlasov-Poisson system with steady states of the Euler-Poisson
system. If $(f,U)$ is a steady state of the
Vlasov-Poisson system with $f(x,v) = \phi (E)$ 
with $\phi$ of the form (\ref{phiass}), then $(\r,p,U)$ satisfy the 
Euler-Poisson system:
Since the solution is static, i.~e.,
\[
u(x) = \frac{1}{\r(x)} \int v f(x,v)\, dv =0,\ x \in \R^3,
\]
the equation $p'=-\r U'$ is all that remains of the Euler equations, and this is
(\ref{ntov}) in the isotropic case $l=0$.
Since the functions $g_m$ in Lemma~\ref{regularity} are strictly decreasing
on their support, $p$ can be written as a function of $\r$, i.~e.,
one obtains the equation of state
\[
p = c_1 g_{1/2} \Bigl(c_2 (g_{-1/2})^{-1}(\r)\Bigr) =
    c_3 \r^{\frac{n+1}{n}} + \O(\r^{\frac{n+1}{n}+\e}),\ \r \to 0+
\]
where $n=k+3/2$, $c_1, c_2, c_3$ are positive constants which depend
on the constants appearing in (\ref{rhodar2}) and (\ref{pdar2}),
and $\e > 0$ depends on $\d$. 
The critical
value $k=3/2$ corresponds to $n=3$, and we now give an example of a 
steady state of the Euler-Poisson system, which has an equation of state
of polytropic form with $3 < n < 5$ for $\r$ small, and with unbounded 
support and infinite mass. It is based on a well-known explicit singular 
solution (cf. \cite{C}, p.\ 89).

\noindent
{\bf Example:}  
Let $n$ be a real number with $3<n<5$ and let ${\cal M}$ be the set of 
functions $\rho(r)$ from $[0,\infty[$ to itself satisfying the following 
conditions:
\begin{enumerate}
\item $\rho$ is a smooth ($C^\infty$) function of $x=r^2$
\item $\rho'(r)<0$ for $r>0$
\item $\rho''(0)<0$
\item $\rho(r)=r^{-\frac{2n}{n-1}}$ for $r>1$
\end{enumerate}
This set of functions is convex. We claim that there is a function $\rho$
in ${\cal M}$ with $\int_0^1 r^2\rho(r) dr=\frac{n-1}{n-3}$. To see this, 
note first that the function $\rho_S (r)=r^{-\frac{2n}{n-1}}$ satisfies this
condition. However, it does not belong to ${\cal M}$. Let $\rho_-$ be an element
of ${\cal M}$ which is everywhere less than or equal to $\rho_S$. It is clear
that such a function exists. It satisfies the condition that
$I_-=\int_0^1 r^2\rho_-(r) dr<\frac{n-1}{n-3}$. It is also clear that 
there exist functions $\rho_+\in{\cal M}$ such that
$I_+=\int_0^1 r^2\rho_+(r) dr$ is as large as desired.
In particular $\rho_+$ can be chosen so that $I_+>\frac{n-1}{n-3}$. There
exists a $\lambda$ in the interval $[0,1]$ such that 
$\lambda I_-+(1-\lambda)I_+=\frac{n-1}{n-3}$. Then
$\rho=\lambda\rho_-+(1-\lambda)\rho_+$ is the desired function.

Define $m(r)= 4\pi\int_0^r s^2 \rho(s) ds$, where $\rho$ is the function just 
constructed. Let $m_S$ be the function constructed in the corresponding
way from $\rho_S$. By construction $\int_0^1 s^2\rho(s)ds=
\int_0^1 s^2\rho_S(s) ds$. Thus $m(1)=m_S(1)$. It follows that $m(r)=m_S(r)$
for all $r>1$. Now define $p(r)=\int_r^\infty s^{-2}m(s)\rho(s)ds$ and $p_S$
analogously. Then for $r>1$ the functions $p$ and $p_S$ are equal. Since
both $\rho$ and $p$ are strictly monotone for $r>0$ it is possible to 
write $p=f(\rho)$ for a smooth function $f$ for $\rho$ in the interval
$(0,\rho(0))$. Moreover, $df/d\rho$ is strictly positive. Both $\rho$ and
$p$ are smooth functions of $x$ for $x\ge 0$. If it could be shown that
$d\rho/dx$ and $dp/dx$ are non-vanishing at $x=0$ then it would follow
that $f$ has a smooth extension to the interval $(0,\infty)$ with everywhere
positive derivative. This is equivalent to showing that $d^2\rho/dr^2$ and 
$d^2 p/dr^2$ are non-vanishing at $r=0$. The first of these two quantities 
is non-vanishing by assumption. The other can easy be computed to be equal 
to $-\frac{4\pi}{3}\rho^2 (0)$. The conclusion is that the functions $\rho$ 
and $p$ are related by an equation of state satisfying all the usual 
conditions, i.e. that for $\rho>0$ the conditions $p>0$ and $dp/d\rho>0$
are satisfied everywhere.

For $r>1$ the function $m(r)$ can be calculated explicitly, as it is equal
to $m_S(r)$. The result is $m(r)=4\pi\frac{n-1}{n-3}r^{\frac{n-3}{n-1}}$.
In the same way $p(r)$ can be calculated explicitly for $r>1$ with the
result $p(r)=4\pi\frac{(n-1)^2}{2(n+1)(n-3)}r^{\frac{-2(n+1)}{n-1}}$
Comparing with the expression for $\rho$ shows that the equation of state
is of polytropic form with index $n$ for $\rho<\rho(1)$.

The example is not dependent on the condition $n<5$. That restriction was 
only made because that is the interesting case where it is known that 
there are {\it some} solutions with finite mass and radius, namely those
with polytropic equation of state. Note also that
it was shown in \cite{RS} that there exist some solutions of the 
Euler-Poisson and Euler-Einstein equations which have an equation of
state which is asymptotically like a polytropic one with $3<n<5$ (but
not exactly polytropic) which have finite radius. This was done by a
perturbation argument with the attendant disadvantages. The above example
serves to show that the strategy applied by Makino\cite{M} cannot be 
directly extended to the case $n>3$ and strongly suggests that the 
methods of this paper cannot be modified to cover cases with 
$k>3/2$. The polytropic steady states are in some sense structurally 
unstable.

The second observation is the following: Consider for $k>0$
the energy-Casimir functional
\[
{\cal D}(f) = 
\frac{k}{k+1}\int\!\!\int f^{1+1/k}\,dv\,dx +{1\over 2} \int\!\!\int v^2 f\,dv\,dx
-\frac{1}{8\pi} \int |\nabla U_f|^2\,dx,
\]
defined on the set
\[
{\cal F}_M = \Bigl\{ f \in L^1 \cap L^{1+1/k} (\R^6) \mid
f \geq 0,\ \int\!\!\int f\, dv\,dx = M,\ 
\int\int v^2 f\,dv\,dx < \infty\Bigr\},
\]
where $U_f$ denotes the potential induced by $f$ and $M>0$ is a prescribed
constant.
One can show that ${\cal D}$ is bounded from below on ${\cal F}_M$
if $k< 3/2$, whereas this functional is not bounded from below on
this set if $k>3/2$. In the first case one can then show that 
the functional actually has a minimizer, at least if one restricts
the set ${\cal F}_M$ to spherically symmetric functions, and
this minimizer is a steady state of the Vlasov-Poisson system of
polytropic form which is dynamically stable in a well-defined sense,
cf.~\cite{G,GR}. Again we see that there seems
to occur a loss of stability if one crosses the threshold $k=3/2$,
i.~e., $n=3$.
In the Casimir part of the functional ${\cal D}$
the expression  $f^{1+1/k}$
can be replaced by more general functions $Q(f)$,
provided $Q$ satisfies growth conditions both for
small and for large values of $f$, which have to satisfy the same
restrictions as $k$ above. This leads to stable steady states
not necessarily of polytropic form.

\end{document}